\documentclass[12pt]{iopart}
\usepackage{epsfig}
\usepackage{graphicx}
\usepackage{amssymb}
\usepackage{bbm}
\usepackage{moreverb}
\usepackage[dvips]{color}
\usepackage[normalem]{ulem}

\usepackage{color}
\definecolor{deepblue}{rgb}{0,0,0.5}
\definecolor{deepred}{rgb}{0.6,0,0}
\definecolor{deepgreen}{rgb}{0,0.5,0}
\DeclareFixedFont{\ttb}{T1}{txtt}{bx}{n}{8} 
\DeclareFixedFont{\ttm}{T1}{txtt}{m}{n}{8}  

\usepackage{listings}
\newcommand\pythonstyle{\lstset{
language=Python,
basicstyle=\ttm,
otherkeywords={self},             
keywordstyle=\ttb\color{deepblue},
emph={def,for,print,in,yield,return,import,from,while,del,if},          
emphstyle=\ttb\color{deepred},    
stringstyle=\color{deepgreen},
frame=tb,                         
showstringspaces=false            %
}}

\lstnewenvironment{python}[1][]
{
\pythonstyle
\lstset{#1}
}
{}

\begin{document}

\newcommand{\binom}[2]{\left(\begin{array}{c} #1\\ #2\end{array}\right)}
\newcommand{\StirOne}[2]{\left[\begin{array}{c} #1\\ #2\end{array}\right]}
\newcommand{\StirTwo}[2]{\left\{\begin{array}{c} #1\\ #2\end{array}\right\}}
\newcommand{\En}[2]{\left\langle \begin{array}{c} #1\\ #2\end{array}\right\rangle}

\title{Matrix product operator representation of polynomial interactions}
\author{Michael L. Wall}
\address{Department of Physics, Colorado School of Mines, Golden, Colorado 80401, USA and The Johns Hopkins University Applied Physics Laboratory, Laurel, MD, 20723, USA}

\begin{abstract}
We provide an exact construction of interaction Hamiltonians on a one-dimensional lattice which grow as a polynomial multiplied by an exponential with the lattice site separation as a matrix product operator (MPO), a type of one-dimensional tensor network. We show that the bond dimension is $(k+3)$ for a polynomial of order $k$, independent of the system size and the number of particles.  Our construction is manifestly translationally invariant, and so may be used in finite- or infinite-size variational matrix product state algorithms.  Our results provide new insight into the correlation structure of many-body quantum operators, and may also be practical in simulations of many-body systems whose interactions are exponentially screened at large distances, but may have complex short-distance structure.
\end{abstract}

\section{Introduction}
\label{sec:Intro}

Starting with the seminal work of Affleck, Kennedy, Lieb, and Tasaki~\cite{Affleck_Kennedy_87}, matrix product states (MPSs), also known as finitely correlated states~\cite{fannes_nachtergaele_92}, have garnered a great deal of theoretical attention.  One of the most appealing features of MPSs is that they provide an exact representation of certain translationally invariant quantum states, and they are unique among state ans\"atze in their ability to do so for entangled states.  In addition to their usefulness as an analytic tool, MPSs are also the underpinning of the density-matrix renormalization group method (DMRG), which has become the \emph{de facto} standard for strongly correlated systems in one spatial dimension (1D).  In particular, DMRG can be expressed as a variational method within the space of MPSs~\cite{Schollwoeck_11}.  The generalization of such a variational ansatz from pure states to density operators led to the introduction of the operator-valued generalization of MPSs, matrix product operators (MPOs), by Verstraete {\it et al.}~\cite{verstraete_garciaripoll_04}.  Later, McCulloch~\cite{Mcculloch_07} realized that significant gains can be had if all operators used in an MPS calculation are represented as MPOs, and put forward a lower triangular ``canonical form" for MPOs.  As an example, the use of MPOs to represent the Hamiltonian operator leads to amortized linear scaling of DMRG sweeps with the system size using caching methods~\cite{Crosswhite_Bacon_08}, and exact arithmetic can be used on MPOs to obtain quantities like the energy variance of quantum states.  Hence, extending the class of operators with exactly known MPO representations not only improves our knowledge of the correlation structure of many-body objects, but also can lead to practical gains in numerical simulations.

Just as MPSs naturally describe quantum states with exponentially decaying correlations~\cite{Schollwoeck_11}, MPOs are most naturally suited to describing interactions which have an exponential decay.  Pirvu et al~\cite{pirvu_murg_10} and Crosswhite and Doherty~\cite{Crosswhite_Doherty_08} showed how general decaying functions may be approximated with MPOs by fitting the functional decay to a sum of MPOs.  As the number of exponentials increases, the interaction is approximated to a larger distance, but for any finite number of exponentials there is a range beyond which the interaction no longer accurately approximates the true function.  In parallel with analytical representations, numerical methods exist also for combining MPOs through arithmetic operations, or for reducing the bond dimension of an MPO~\cite{PhysRevB.95.035129}.  In Ref.~\cite{Froewis_Nebendahl_10}, Fr\"owis, Nebendahl, and D\"ur undertook a classification of Hamiltonians which have an MPO representation whose bond dimension is independent of the system size.  One of their examples was a polynomial times an exponential function, which was claimed to have a bond dimension of $\mathcal{O}\left(k\right)$, with $k$ being the order of the polynomial, independent of the system size.  An example MPO was given, but no constructive method nor proof of the methodology was presented for general polynomials.  In this paper, we put forth a constructive characterization of the MPO representation of Hamiltonians with polynomial times exponential interactions for all orders $k$, and show that the bond dimension is $(k+3)$.

The organization is as follows: In Sec.~\ref{sec:MPOIntro} we briefly review the theory of MPOs to set notation and discuss previously known examples.  In Sec.~\ref{sec:PPLI} we present an MPO ansatz for general positive power-law interactions and prove its validity.  Sec.~\ref{sec:GPI} generalizes the results of the previous section to general polynomial interactions.  Finally, in Sec.~\ref{sec:concl} we conclude and give an outlook.  Python code to solve for the coefficients of the MPO ansatz and a table of these coefficients in the special cases of power law interactions for the first six powers are given as appendices.

\section{Matrix product operator definitions and examples}
\label{sec:MPOIntro}

Let us consider a lattice of $L$ sites, each of which contains a $d$-dimensional Hilbert space spanned by the states $\{|i\rangle, i=1,\dots,d\}$.  A matrix product operator (MPO) acting on the Hilbert space of this lattice is defined as
\begin{eqnarray}
\label{eq:MPOdef}\hat{O}&=&\sum_{i_1,\dots i_L, i_1', \dots, i_L'}\mathrm{Tr}\left[\mathbb{W}^{i_1i_1'\left[1\right]}\dots \mathbb{W}^{i_Li_L'\left[1\right]}\right]|i_1\dots i_L\rangle \langle i_1'\dots i_L'|\, ,
\end{eqnarray}
where each of the objects $\mathbb{W}^{i_ji_j'\left[j\right]}$ is a matrix whose linear dimension is bounded by $\chi$, which we call the bond dimension of the matrix product operator, and $\mathrm{Tr}$ denotes the matrix trace.  The indices of the physical Hilbert space, e.g., $i_j$ are called physical indices, while those involved in the matrix product and trace will be referred to as bond indices.  It is useful to re-write this expression as 
\begin{eqnarray}
\label{eq:MPOdefmathcal}\hat{O}&=&\mathrm{Tr}\left[\hat{\mathcal{W}}^{\left[1\right]}\dots \hat{\mathcal{W}}^{\left[L\right]}\right]\, ,
\end{eqnarray}
where now each of the $\hat{\mathcal{W}}^{\left[j\right]}\equiv\sum_{i_ji_j'}\mathbb{W}^{i_ji_j'\left[j\right]}|i_j\rangle\langle i_j|$ is a matrix of operators acting on the Hilbert spaces spanned by the $\{|i_j\rangle\}$ and whose matrix indices are the same as the $\mathbb{W}^{i_ji_j'\left[j\right]}$.

For a translationally invariant system, only a single MPO matrix $\hat{\mathcal{W}}$ suffices to describe the operator.  If we are constructing the representation of this operator on a finite chain of $L$ sites with open boundary conditions, as is the most common scenario for numerical MPS simulations, we simply take the first MPO matrix $\hat{\mathcal{W}}^{[1]}$ to be the last row of $\hat{\mathcal{W}}$, the last MPO matrix $\hat{\mathcal{W}}^{[L]}$ to be the first column of $\hat{\mathcal{W}}$, and all other MPO matrices $\hat{\mathcal{W}}^{[1<j<L]}$ to be $\hat{\mathcal{W}}$.  In what follows, we will focus on such translationally invariant operators, and hence only describe the single MPO operator $\hat{\mathcal{W}}$.

Similar to MPS representations of quantum states, MPOs are remarkable in their ability to compactly represent many-body operators.  As an example, the MPO matrix $\hat{\mathcal{W}}_{\mathrm{one-body}}$ describing a one-body operator $\sum_i\hat{X}_i$ is
\begin{eqnarray}
\hat{\mathcal{W}}_{\mathrm{one-body}}&=&\left(\begin{array}{cc} \hat{I}&0 \\ \hat{X}&\hat{I}\end{array}\right)\, ,
\end{eqnarray}
the matrix $\hat{\mathcal{W}}_{\mathrm{two-body}}$ for a two-body operator $\sum_{i}\hat{X}_i\hat{Y}_{i+1}$ is 
\begin{eqnarray}
\hat{\mathcal{W}}_{\mathrm{two-body}}&=&\left(\begin{array}{ccc} \hat{I}&0&0 \\ \hat{Y}&0&0 \\ 0&\hat{X}&\hat{I}\end{array}\right)\, ,
\end{eqnarray}
and that for an exponentially decaying interaction $\sum_{i<j}\beta^{j-i}\hat{X}_i\hat{Y}_j$ is
\begin{eqnarray}
\label{eq:Wexp}\hat{\mathcal{W}}_{\mathrm{exponential}}&=&\left(\begin{array}{ccc} \hat{I}&0&0 \\ \hat{Y}&\beta\hat{I}&0 \\ 0&\beta\hat{X}&\hat{I}\end{array}\right)\, .
\end{eqnarray}
In all these examples, the dimensions of the matrices are indexed by bond indices, while the physical Hilbert space is described by the operator character of, e.g., $\hat{X}$.  More examples can be found in the literature, e.g., Ref.~\cite{Froewis_Nebendahl_10}.

\section{MPO construction of positive power-law interactions}
\label{sec:PPLI}

\subsection{Statement of MPO ansatz}
Our ansatz for the MPO matrix describing the Hamiltonian of a power-law interaction of the form
\begin{eqnarray}
\label{eq:powerHami}\hat{H}&=&\sum_{i<j}\left(j-i\right)^k\hat{X}_i\hat{Y}_j\, ,\;\;k\in\mathbb{N}\, ,
\end{eqnarray}
is
\begin{eqnarray}
\label{eq:MPOmatrix} \hat{\mathcal{W}}&=&\left(\begin{array}{ccccccc}
\hat{I}&0&0&\dots&0&0&0\\
\frac{\hat{Y}}{\sqrt{k+1}}&\hat{I}&0&\dots&0&0&0\\
\frac{\hat{Y}}{\sqrt{k+1}}&a_{1}\hat{I}&\hat{I}&\dots&0&0&0\\
\vdots&\vdots&\ddots&\ddots&\vdots&\vdots&\vdots\\
\frac{\hat{Y}}{\sqrt{k+1}}&a_{k-1}\hat{I}&a_{k-2}\hat{I}&\dots &\hat{I}&0&0\\
\frac{\hat{Y}}{\sqrt{k+1}}&a_{k}\hat{I}&a_{k-1}\hat{I}&\dots &a_{1}\hat{I}&\hat{I}&0\\
0&\frac{\hat{X}}{\sqrt{k+1}}&\frac{\hat{X}}{\sqrt{k+1}}&\dots &\frac{\hat{X}}{\sqrt{k+1}}&\frac{\hat{X}}{\sqrt{k+1}}&\hat{I}\\
\end{array}\right)\, ,
\end{eqnarray}
or, in a more compact notation,
\begin{eqnarray}
\label{eq:ansatz}\hat{\mathcal{W}}&=&\left(\begin{array}{ccc}
\hat{I}&0&0\\
\frac{\hat{Y}}{\sqrt{k+1}}\mathbf{1}^T_{k+1}&\mathbb{L}_{k}\left(\mathbf{a}\right)\hat{I}&0\\
0&\frac{\hat{X}}{\sqrt{k+1}}\mathbf{1}_{k+1}&\hat{I} \end{array}\right)\, .
\end{eqnarray}
In Eq.~\eref{eq:ansatz}, $\mathbf{1}_n$ is a vector of length $n$ whose elements are all 1 and $\mathbb{L}_k\left(\mathbf{a}\right)$ is the $(k+1)\times (k+1)$ matrix
\begin{eqnarray}
\mathbb{L}_k\left(\mathbf{a}\right)&=\left(\begin{array}{ccccccc}
1		&		0		&		0		&		\dots			&		0			&		0		&		0\\
a_{1}      &		1		&		0		&		\dots			&		0			&		0		&		0\\
a_{2}      &		a_{1}	&		1		&		\dots 		&		0			&		0		&		0\\
\vdots	&		\ddots	&		\ddots	&		\ddots		&		\vdots		&		\vdots	&		\vdots\\
a_{k-2}		&		a_{k-3}		&		\ddots		&		\ddots		&		1			&		0		&		0\\
a_{k-1}		&		a_{k-2}		&		a_{k-3}		&		\ddots		&		a_{1}		&		1		&		0\\
a_k		&		a_{k-1}		&		a_{k-2}		&		\dots			&		a_{2}		&		a_{1}		&		1\\
\end{array}\right)\, ,
\end{eqnarray}
parameterized by the vector $\mathbf{a}=\left(a_1,a_1,\dots a_{k}\right)$.  Note that the indices of the matrix $\mathbb{L}_k$ and the vectors $\mathbf{1}_{k+1}$ are bond indices.  As in the above, only operators denoted by hats act on the physical Hilbert space.  $\mathbb{L}_k\left(\mathbf{a}\right)$ is a Toeplitz matrix, i.e.~a diagonally constant matrix, and is lower triangular.  Comparing with the exponential MPO in Eq.~\eref{eq:Wexp}, we see that a polynomial multiplied by an exponential interaction of the form
\begin{eqnarray}
\hat{H}&=&\sum_{i<j}\beta^{j-i}\left(j-i\right)^k\hat{X}_i\hat{Y}_j\, ,
\end{eqnarray}
is immediately obtained by replacing $\mathbb{L}_{k}\left(\mathbf{a}\right)\to \beta \mathbb{L}_{k}\left(\mathbf{a}\right)$, $\hat{X}\to\beta\hat{X}$, which does not change the structure or bond dimension of the MPO.

To begin to understand how the ansatz Eq.~\eref{eq:ansatz} generates polynomial interactions, let us consider the Hamiltonian on an increasing number of sites.  For $L=2$ sites, the MPO matrices are $\hat{\mathcal{W}^{[1]}}=\hat{\mathcal{W}}_{k+1,:}$ and $\hat{\mathcal{W}}^{[L]}=\hat{\mathcal{W}}_{:,1}$, as described above.  Hence,
\begin{eqnarray}
\hat{H}_{L=2}&=&\frac{1}{k+1}\left(\mathbf{1}_{k+1}^{T} \mathbf{1}_{k+1}\right)\hat{X}_1\hat{Y}_2=\left(2-1\right)^k\hat{X}_1\hat{Y}_2\, .
\end{eqnarray}
For $L=3$ sites, $\hat{\mathcal{W}}^{[1]}$ and $\hat{\mathcal{W}}^{[L]}$ remain the same and we introduce an additional matrix $\hat{\mathcal{W}}^{[L-1]}=\hat{\mathcal{W}}$ in between them, finding
\begin{eqnarray}
\fl\hat{H}_{L=3}&=&\frac{1}{k+1}\left(\mathbf{1}_{k+1}^{T} \mathbf{1}_{k+1}\right)\left[\hat{X}_1\hat{Y}_2+\hat{X}_2\hat{Y}_3\right]+\frac{1}{k+1}\left(\mathbf{1}_{k+1}^{T} \mathbb{L}_k\left(\mathbf{a}\right)\mathbf{1}_{k+1}\right)\hat{X}_1\hat{Y}_3\, .
\end{eqnarray}
This gives the condition that $\mathbf{1}^{T}_{k+1}\mathbb{L}_k\left(\mathbf{a}\right) \mathbf{1}_{k+1}=\left(k+1\right)2^k$ for our ansatz to faithfully reproduce the power-law interaction at this length.  Following this line of reasoning through, the conditions on the vector $\mathbf{a}$ such that the MPO matrix Eq.~\eref{eq:MPOmatrix} reproduces the Hamiltonian Eq.~\eref{eq:powerHami} are
\begin{eqnarray}
\label{eq:kequations} \mathbf{1}^{T}_{k+1}\mathbb{L}_k^n\left(\mathbf{a}\right) \mathbf{1}_{k+1}&=&\left(k+1\right)\left(n+1\right)^k\,  
\end{eqnarray}
for $n=1,\dots, k$; that is, conditions are placed on the elementwise sums of powers of the $\mathbb{L}_k\left(\mathbf{a}\right)$ matrix.  An inductive proof that this set of equations produces the Hamiltonian on any number of lattice sites is saved for Sec.~\ref{sec:Proof}.  Eq.~\eref{eq:kequations} represents a system of $k$ equations in $k$ unknowns.  However, the $n^{\mathrm{th}}$ equation is a degree $n$ polynomial in products of the elements of $\mathbf{a}$, and the solution of Eq.~\eref{eq:kequations} is hence a nontrivial task.

\subsection{Formulation of the constraint equations}
\label{sec:form}
We begin the solution of Eq.~\eref{eq:kequations} by defining the $n\times n$ shift matrix, $\mathbb{Z}_n$, as
\begin{eqnarray}
\mathbb{Z}_n&\equiv& \left(\begin{array}{ccccc} 0&0&\dots&0&0\\ 1&0&\dots&0&0\\ 0&1&\dots&0&0\\ \vdots&\vdots&\ddots&\vdots&\vdots\\ 0&0&\dots&1&0\end{array}\right)\, .
\end{eqnarray}
From its definition, $\mathbb{Z}_n$ is nilpotent with degree $n$, $\mathbb{Z}_n^n=0$.  We can write $\mathbb{L}_k\left(\mathbf{a}\right)$ in terms of the shift matrix as
\begin{eqnarray}
\label{eq:LinZexpansion} \mathbb{L}_k\left(\mathbf{a}\right)&=&\sum_{i=0}^{k}a_{i}\mathbb{Z}_{k+1}^i\, ,
\end{eqnarray}
where we have set $a_{0}=1$.  Using Eq.~\eref{eq:LinZexpansion}, the $n^{\mathrm{th}}$ power of $L_{k}\left(\mathbf{a}\right)$ also has a power series expansion in $\mathbb{Z}_{k+1}$
\begin{eqnarray}
\mathbb{L}_k^n\left(\mathbf{a}\right)&=&\left(\sum_{j=0}^{k}a_{j}\mathbb{Z}_{k+1}^j\right)^n=\sum_{j=0}^{k}c_j^{\left(n,k\right)}\mathbb{Z}_{k+1}^j\, ,
\end{eqnarray}
where the coefficients $c_j^{\left(n,k\right)}$ are defined recursively as
\begin{eqnarray}
\label{eq:crecursion}c_0^{\left(n,k\right)}\equiv 1\, ,\;\; c_m^{\left(n,k\right)}&=&\frac{1}{m}\sum_{j=1}^{m}\left[j\left(n+1\right)-m\right]a_{j}c_{m-j}^{\left(n,k\right)}\, .
\end{eqnarray}
Furthermore, $\mathbf{1}_n^T\cdot A\cdot \mathbf{1}_n=\sum_{ij=1}^nA_{ij}$ for any $n\times n$ matrix $\mathbb{A}$, and so the constraint equations Eq.~\eref{eq:kequations} may be stated in terms of the coefficients $c_j^{\left(n,k\right)}$ as
\begin{eqnarray}
\label{eq:basic}\xi_{nk}\equiv \sum_{j=1}^{k}c_j^{\left(n,k\right)}\left(k+1-j\right)&=&\left(k+1\right)\left[\left(n+1\right)^k-1\right]\, .
\end{eqnarray}

Let us now derive a recursion relation between the coefficients $c_j^{\left(n,k\right)}$ with different $n$.  We do so by equating powers of $\mathbb{Z}_{k+1}$ in the expansion
\begin{eqnarray}
\fl\mathbb{L}_k^n\left(\mathbf{a}\right)&=&\sum_{j=0}^{k}c_j^{\left(n,k\right)}\mathbb{Z}_{k+1}^j=\mathbb{L}_k^{n-1}\mathbb{L}_k=\left(\sum_{j=0}^{k}c_j^{\left(1,k\right)}\mathbb{Z}_{k+1}^{j}\right)\left(\sum_{j=0}^{k}c_j^{\left(n-1,k\right)}\mathbb{Z}_{k+1}^j\right)\, ,
\end{eqnarray}
and find
\begin{eqnarray}
\label{eq:ncrecursion} c_j^{\left(nk\right)}&=&c_j^{\left(n-1,k\right)}+\sum_{p=1}^{j}c_{p}^{\left(1,k\right)}c_{j-p}^{\left(n-1,k\right)}\, .
\end{eqnarray}
Repeatedly applying the recursion Eq.~\eref{eq:ncrecursion} on the right hand side of Eq.~\eref{eq:ncrecursion}, we find
\begin{eqnarray}
c_j^{\left(n,k\right)}&=&c_j^{\left(0,k\right)}+\sum_{q=1}^{n}\binom{n}{q}\sum_{p_1\dots p_q}'c_{j-\sum_ip_i}^{\left(0,k\right)}\prod_ic_{p_i}^{\left(1,k\right)}\, .
\end{eqnarray}
Here, the primed summation is defined as
\begin{eqnarray}
\sum_{p_1\dots p_q}'&\equiv&\sum_{p_1=1}^{j}\sum_{p_2=1}^{p_1}\dots \sum_{p_q=1}^{p_{q-1}}\, .
\end{eqnarray}
Noting that $c_j^{\left(0,k\right)}=\delta_{j,0}$ and $c_j^{\left(1,k\right)}=a_{j}$, we find
\begin{eqnarray}
c_j^{\left(n,k\right)}&=&\sum_{q=1}^{n}\binom{n}{q}\sum_{p_1+\dots+p_q=j}a_{p_1}\dots a_{p_q}\, ,
\end{eqnarray}
where $p_i\ge 1$ and $j\ge 1$.  Stated in terms of $\xi_{nk}$, we have
\begin{eqnarray}
\label{eq:xiRestatement} \xi_{nk}&=\sum_{j=1}^{k}\left(k+1-j\right)\sum_{q=1}^{n}\binom{n}{q} \sum_{p_1+\dots+p_q=j}a_{p_1}\dots a_{p_q}\, .
\end{eqnarray}

The result Eq.~\eref{eq:xiRestatement} is a precise restatement of the fact that the $n^{\mathrm{th}}$ order condition Eq.~\eref{eq:kequations} is a degree-$n$ polynomial in products of the elements of $\mathbf{a}$.  In order to simplify the equations it is convenient to work not directly with the sequence $\xi_{nk}$, but with its binomial transform
\begin{eqnarray}
\label{eq:etadef} \sum_{j=1}^{m}\binom{m}{j}\eta_{jk}&=&\xi_{mk}\, ,\;\; \eta_{mk}=\sum_{j=1}^{m}\left(-1\right)^{j+m}\xi_{jk}\binom{m}{j}\, .
\end{eqnarray}
We take the sums from $j=1$ due to the fact that $\xi_{0k}=\eta_{0k}=0$.  Using Eqs.~\eref{eq:etadef} and \eref{eq:xiRestatement} together, we have that
\begin{eqnarray}
\fl\eta_{mk}&=&\sum_{n=1}^{m}\left(-1\right)^{n+m}\binom{m}{n}\sum_{j=1}^{k}\left(k+1-j\right)\sum_{q=1}^{n}\binom{n}{q} \sum_{p_1+\dots+p_q=j}a_{p_1}\dots a_{p_q}\\
\fl&=&\sum_{j=1}^{k}\left(k+1-j\right)\sum_{p_1+\dots+p_m=j}a_{p_1}\dots a_{p_m}\, .
\end{eqnarray}
Because of the restriction that all indices $p_i\ge 1$, the condition $\sum_{i=1}^{m}p_i=j$ can only be satisfied for $j\ge m$, and so
\begin{eqnarray}
\label{eq:etaa}\eta_{mk}&=\sum_{j=m}^{k}\left(k+1-j\right)\sum_{p_1+\dots+p_m=j}a_{p_1}\dots a_{p_m}\, .
\end{eqnarray}
That is, each term in $\eta_{mk}$ is a monomial of degree $m$ in the elements of $\mathbf{a}$.  Furthermore, from the sum restriction, $\eta_{mk}$ involves only the elements $a_{p}$ with $p\le k-m+1$.  In particular, we have that
\begin{eqnarray}
\eta_{kk}&=a_{1}^k\, .
\end{eqnarray}
Hence, we may solve for $a_1$, $a_2$, etc.~in ascending order by considering the expressions Eq.~\eref{eq:etaa} in descending order of $m$.  A numeric value for $\eta_{mk}$ is obtained by using the far right-hand side of Eq.~\eref{eq:basic}, and yields
\begin{eqnarray}
\label{eq:etaNumeric}\eta_{mk}=\left(k+1\right)\sum_{j=1}^{m}\left(-1\right)^{j+m}\binom{m}{j}\left[\left(j+1\right)^k-1\right]\, .
\end{eqnarray}
Expanding the power on the right hand side using the binomial theorem and applying the definition of the Stirling numbers of the second kind
\begin{eqnarray}
\StirTwo{p}{k}&\equiv&\frac{1}{k!}\sum_{j=1}^{k}\left(-1\right)^{k-j}\binom{k}{j}j^p\, ,
\end{eqnarray}
we find
\begin{eqnarray}
\eta_{mk}=\left(k+1\right)m!\sum_{q=1}^{k}\binom{k}{q}\StirTwo{q}{m}\, .
\end{eqnarray}
Using the identity~\cite{graham1989concrete}
\begin{eqnarray}
\sum_{p}\binom{n}{p}\StirTwo{p}{m}&=\StirTwo{n+1}{m+1},
\end{eqnarray}
we then have
\begin{eqnarray}
\eta_{mk}=\left(k+1\right)m!\StirTwo{k+1}{m+1}\, .
\end{eqnarray}
In particular, for $m=k$, we have $\eta_{kk}=(k+1)!$.  Hence, if we choose the positive real root $a_{1}=\left[\left(k+1\right)!\right]^{1/k}$, then the entire vector $\mathbf{a}$ may be taken to be real.  As an example, the equations to be solved for $k=4$ are
\begin{eqnarray}
\label{eq:eta44}\eta_{44}&&=a_1^4=120\\
\eta_{34}&&=2a_1^3+3a_1^2a_2=300\\
\eta_{24}&&=3a_1^2+4a_1a_2+2a_1a_3+a_2^2=250\\
\label{eq:eta14}\eta_{14}&&=4a_1+3a_2+2a_3+a_4=75\, .
\end{eqnarray}

\subsection{Solution of the constraint equations}
\label{sec:soln}

While $a_1$ may be found analytically, the other coefficients must be generated numerically.  To derive an efficient numerical procedure, we return to Eq.~\eref{eq:etaa} and note that
\begin{eqnarray}
\label{eq:etaLHS}\eta_{mk}&=\left(k+1-m\right)a_{k-1}^m+\sum_{q=1}^{k-m}\left(k+1-m-q\right)P_{mq}^{k}
\end{eqnarray}
where
\begin{eqnarray}
P_{mq}^k&=\sum_{p_1+\dots+p_m=m+q}a_{k-p_1}\dots a_{k-p_m}\, .
\end{eqnarray}
The sum counts the number of ways to partition the integer $(m+q)$ into $m$ pieces such that each piece $p_i\ge 1$.  Equivalently, the problem is the number of integer partitions of $q$ into at most $m$ pieces.  This is a standard problem in combinatorics, and will not be reviewed here; an implementation is given as part of the python program in \ref{sec:AppPython}.   A particular partition $p$ may be written as the set $\left\{\mathbf{s},\mathbf{m}\right\}$, where $\mathbf{s}$ denotes the distinct integers forming the partition and $\mathbf{m}$ denotes their \emph{multiplicities}, the number of times each integer appears, such that $\sum_is_im_i=n$.  We will refer to the number of distinct integers in a particular partition $p$ as the length of the partition, and denote it with $\ell$.  In terms of these quantities, the number of different ways that a particular partition $p$ may be realized is $\Omega\left(p\right)={\left(\sum_im_i\right)_{\ell}}/{\prod_im_i!}$.  The numerator counts the number of ways of arranging the $p_i=1$, and the denominator removes identical rearrangements of the other $p_j$.  With all of the integer partitions and their multiplicities, we can generate the monomials in Eq.~\eref{eq:etaLHS} and their weights.  Substituting the numerical values of previously solved components of the vector $\mathbf{a}$, this becomes an equation for a single unknown component $a_j$, see, e.g.~\eref{eq:eta44}-\eref{eq:eta14}.  The numerical right hand side of the equation is obtained from Eq.~\eref{eq:etaNumeric}.  Hence, starting from $a_1$, the entire vector $\mathbf{a}$ can be obtained to any desired numerical precision.  A python implementation of this procedure is given as \ref{sec:AppPython}.

\subsection{Proof of construction}
\label{sec:Proof}
In this section we prove that if the conditions Eq.~\eref{eq:kequations} hold for $n=1,\dots, k$, then they hold for any $n\in\mathbb{N}$.  This demonstrates that the MPO matrix Eq.~\eref{eq:MPOmatrix} faithfully represents the Hamiltonian Eq.~\eref{eq:powerHami} for a system of any number of sites.  The proof is inductive.  Let us assume that Eq.~\eref{eq:kequations} is true for all $n=1,\dots, k$, and venture to prove that Eq.~\eref{eq:kequations} for $n=(k+1)$ follows, that is,
\begin{eqnarray}
\mathbf{1}^{T}_{k+1}\mathbb{L}_k^{k+1}\left(\mathbf{a}\right) \mathbf{1}_{k+1}&=&\left(k+1\right)\left(k+2\right)^k\, .
\end{eqnarray}
We begin by noting that the eigenvalues of $\mathbb{L}_k\left(\mathbf{a}\right)$ are its diagonal elements, as is true for any triangular matrix.  Hence, $\mathbb{L}_k\left(\mathbf{a}\right)$ satisfies the characteristic polynomial equation
\begin{eqnarray}
\label{eq:chareqn} \left(\mathbb{L}_k\left(\mathbf{a}\right)-\mathbb{I}\right)^{k+1}&=&0\, .
\end{eqnarray} 
Using the binomial theorem and rearranging, we find
\begin{eqnarray}
\mathbb{L}_k^{k+1}\left(\mathbf{a}\right)&=&\sum_{r=0}^{k}\binom{k+1}{r}\left(-1\right)^{k-r}\mathbb{L}_k^r\left(\mathbf{a}\right)\, .
\end{eqnarray}
Multiplying by $\mathbf{1}_{k+1}$ on the right, by $\mathbf{1}_{k+1}^T$ on the left, and using the hypotheses $\mathbf{1}_{k+1}^T\mathbb{L}_k^{n}\left(\mathbf{a}\right)\mathbf{1}_{k+1}=\left(k+1\right)\left[\left(n+1\right)^k-1\right]$, $n=1,\dots, k$, we find
\begin{eqnarray}
\mathbf{1}_{k+1}^T\mathbb{L}_k^{k+1}\left(\mathbf{a}\right)\mathbf{1}_{k+1}&=&\left(k+1\right)\sum_{r=1}^{k}\binom{k+1}{r}\left(-1\right)^{k-r}\left(r+1\right)^k\, .
\end{eqnarray}
By virtue of Worpitzky's identity,
\begin{eqnarray}
x^k&=&\sum_{q=0}^{k-1}\binom{q+x}{k}\En{k}{q}\, ,
\end{eqnarray}
with $\En{k}{q}$ an Eulerian number, we have
\begin{eqnarray}
\label{eq:uphere}\fl \mathbf{1}_{k+1}^T\mathbb{L}_k^{k+1}\left(\mathbf{a}\right)\mathbf{1}_{k+1}&=&\left(k+1\right)\sum_{q=0}^{k-1}\left[\sum_{r=1}^{k}\left(-1\right)^{k-r}\binom{k+1}{r}\binom{q+r+1}{k}\right]\En{k}{q}\, .
\end{eqnarray}
To determine the value of the term in brackets, we write
\begin{eqnarray}
\fl\sum_{r=1}^{k+1}\left(-1\right)^{k-r}\binom{k+1}{r}\binom{q+r+1}{k}&=&\sum_{r=1}^{k+1}\left(-1\right)^{k-r}\binom{k+1}{r}\frac{\left(q+r+1\right)_k}{k!}\, ,
\end{eqnarray}
where $\left(x\right)_k$ is the falling factorial.  Expanding the falling factorial in terms of the Stirling numbers of the first kind as
\begin{eqnarray}
\left(x\right)_n&=&\sum_{p=0}^{n}\left(-1\right)^{n-p}\StirOne{n}{p}x^p\, ,
\end{eqnarray}
we have
\begin{eqnarray}
\nonumber \fl&&\sum_{r=1}^{k+1}\left(-1\right)^{k-r}\binom{k+1}{r}\binom{q+r+1}{k}\\
\label{eq:inbrackets}\fl &=&\frac{1}{k!}\sum_{p=0}^{k}\left(-1\right)^{p}\StirOne{k}{p}\sum_{\ell=0}^{p}\binom{p}{\ell}q^{p-\ell}\left[\sum_{j=0}^{k+1}\binom{k+1}{j}j^{\ell}\left(-1\right)^j\right]\, .
\end{eqnarray}
The term in brackets in Eq.~\eref{eq:inbrackets} vanishes according to the identity
\begin{eqnarray}
\sum_{i=0}^{k}\left(-1\right)^{i}\binom{k}{i}i^m&=&0\, ,\;\; 0\le m\le k\, ,\;\;m,k\in\mathbb{Z}\, ,
\end{eqnarray}
which may be proved inductively.  Hence,
\begin{eqnarray}
\sum_{r=1}^{k}\left(-1\right)^{k-r}\binom{k+1}{r}\binom{q+r}{k}&=\binom{k+q+1}{k}\, ,
\end{eqnarray}
and so Eq.~\eref{eq:uphere} becomes
\begin{eqnarray}
\mathbf{1}_{k+1}^T\mathbb{L}_k^{k+1}\left(\mathbf{a}\right)\mathbf{1}_{k+1}&=&\left(k+1\right)\sum_{q=1}^{k}\binom{k+q+1}{k}\En{k}{q}\, .
\end{eqnarray}
Using Worpitzky's identity again, we have
\begin{eqnarray}
\mathbf{1}_{k+1}^T\mathbb{L}_k^{k+1}\left(\mathbf{a}\right)\mathbf{1}_{k+1}&=&\left(k+1\right)\left(k+2\right)^k\, ,
\end{eqnarray}
as was to be shown.  This proof may be generalized to any arbitrary integer $n=k+\ell$, $\ell\in\mathbb{N}$, by multiplying the characteristic polynomial Eq.~\eref{eq:chareqn} by $\left(\mathbb{L}_k\left(\mathbf{a}\right)-I\right)^{\ell-1}$ and following an identical line of reasoning.  Hence, once the values of $\mathbf{a}$ have been set by the conditions of Eq.~\eref{eq:kequations} for $n=1,\dots,k$, the MPO matrix Eq.~\eref{eq:MPOmatrix} reproduces the Hamiltonian Eq.~\eref{eq:powerHami} on any number of sites.

\section{Extension to general polynomial interactions}
\label{sec:GPI}
The proof given in Sec.~\ref{sec:Proof} relies only on the form of Worpitzky's identity
\begin{eqnarray}
x^k&=&\sum_{q=0}^{k-1}\binom{q+x}{k}\En{k}{q}\, .
\end{eqnarray}
Hence, the same analysis applies to any function $P_k\left(x\right)$ which can be written as a linear combination of $\left\{\binom{q+x}{k}\right\}$, $q=0,\dots,k-1$.  Because these binomial coefficients form a basis for the space of polynomials of degree $k$ with no constant term, the most general functions which are linear combinations of these binomial coefficients are degree-$k$ polynomials of the form
\begin{eqnarray}
\label{eq:Polydef}P_k\left(x\right)&=&\sum_{i=1}^{k}\alpha_i x^i\, .
\end{eqnarray}
We now wish to express polynomials of the form Eq.~\eref{eq:Polydef} as
\begin{eqnarray}
\label{eq:degkpoly}P_k\left(x\right)&=&\sum_{q=0}^{k-1}\binom{q+x}{k}W_{kq}\, .
\end{eqnarray}
We do so by noting that finding the coefficients $W_{kq}$ is equivalent to solving the linear system of equations
\begin{eqnarray}
\fl \left(\begin{array}{cccc} \binom{1+0}{k}&\binom{1+1}{k}&\dots&\binom{1+k-1}{k}\\ \binom{2+0}{k}&\binom{2+1}{k}&\dots&\binom{2+k-1}{k}\\ \vdots&\vdots&\ddots&\vdots\\ \binom{k+0}{k}&\binom{k+1}{k}&\dots&\binom{k+k-1}{k}\end{array}\right)\left(\begin{array}{c}W_{k0}\\ W_{k1}\\ \vdots\\ W_{k,n-1}\end{array}\right)&=&\left(\begin{array}{c} P_k\left(1\right)\\ P_k\left(2\right)\\ \vdots \\ P_k\left(k\right)\end{array}\right)\, .
\end{eqnarray}
This linear system is solved for any $P_k\left(x\right)$ by inverting the Hankel matrix with elements $H^{\left(k\right)}_{ij}=\binom{i+j-1}{k}$.  It can be verified that $\left[H^{\left(k\right)}\right]^{-1}$ is again a Hankel matrix defined by the elements $\left[H^{\left(k\right)}\right]^{-1}_{ij}=\left(-1\right)^{k+1-\left(i+j\right)}\binom{k+1}{k+1-\left(i+j\right)}.$  For the special case $P_k\left(x\right)=x^k$, this construction reproduces the known representation of the Eulerian numbers
\begin{eqnarray}
\En{n}{m}&=&\sum_{k=0}^{m}\left(-1\right)^k\binom{n+1}{k}\left(m+1-k\right)^n\, ,
\end{eqnarray}
and hence Worpitzky's identity.

The above construction demonstrates that for any degree-$k$ polynomial $P_k\left(x\right)$ of the form Eq.~\eref{eq:degkpoly} an exact MPO representation with bond dimension $(k+3)$ may be found.  To find the vector of coefficients $\mathbf{a}$ which reproduces this polynomial, one uses the machinery of Secs.~\ref{sec:form}-\ref{sec:soln} with Eq.~\eref{eq:etaNumeric} replaced by
\begin{eqnarray}
\eta_{mk}=\left(k+1\right)\sum_{j=1}^{m}\left(-1\right)^{j+m}\binom{m}{j}\left[P_k\left(j+1\right)-1\right]\, .
\end{eqnarray}
Also, as noted above, the MPO resulting from this construction is immediately generalized to interactions consisting of polynomials multiplied by an exponential using the replacements $\mathbb{L}_{k}\left(\mathbf{a}\right)\to \beta \mathbb{L}_{k}\left(\mathbf{a}\right)$, $\hat{X}\to\beta\hat{X}$, which does not change the structure or bond dimension of the MPO.

\section{Conclusions}
\label{sec:concl}

To summarize, we have put forwards an exact construction of Hamiltonians consisting of interactions whose strength varies as a degree-$k$ polynomial multiplied by an exponential with site separation as a matrix product operator (MPO) with bond dimension $(k+3)$, independent of the system size or the number of particles.  In addition to a proof that this construction reproduces the desired Hamiltonian on any number of sites, we also described an algorithm to determine the parameters appearing in the MPO ansatz to any desired numerical precision; a python implementation of this algorithm is given as \ref{sec:AppPython}.  In addition to being useful for constructing complex operators for use in variational MPS calculations, our results provide new analytic insight into efficiently constructing quantum states and operators with complex correlations.  

\newpage

\appendix

\section{Table of values of $\mathbf{a}$ for the first six powers}
\label{sec:AppTable}
In table~\ref{tab:tabA}, we collect the numerical values of the vector $\mathbf{a}$ for the first six powers.  These values were generated using the code provided in \ref{sec:AppPython}.

\begin{table}
\tiny
\begin{tabular}{|c|c|c|c|c|c|c|}
\hline $k$&$a_1$&$a_2$&$a_3$&$a_4$&$a_5$&$a_6$\\
\hline 1&2.0000000000000000&&&&&\\
\hline 2&4.1010205144336442&2.4494897427831779&&&&\\
\hline 3&8.4748302749699516&5.4358361515927998&2.8844991406148166&&&\\
\hline 4&17.55558915612346&11.719390564662234&6.9222086786548589&3.3097509196468731&&\\
\hline 5&36.379219139956668&24.929714047082978&15.628482874669601&8.5590769439779582&3.7279192731913513&\\
\hline 6&75.3472962465863&52.61301470336558&34.27556525233372&20.25723076990663&10.345394101852634&4.1406808334652885\\
\hline 
\end{tabular}
\caption{\label{tab:tabA} Table of values of $\mathbf{a}$ for the first six powers}
\end{table}

\section{Python code for solving for $\mathbf{a}$}
\label{sec:AppPython}

\begin{python}
from scipy.misc import comb
from math import factorial, log, exp
    
def Partitions(n,k):
    """Generate all partitions of an integer n into at most k positive integers.
    The partitions are returned as a dict mapping the integer n_i to its multiplicity m_i such that
    \sum_i n_i m_i = n
    """
    if n == 0:
        yield {}
        return    
    partition = {n : 1} #start with trivial partition of n into 1 n
    my_keys = [n] #keys in the partition dict, sorted largest to smallest
    yield partition

    while my_keys != [1]: #work your way down to n ones        
        reuse = 0
        if my_keys[-1] == 1: #If my last generated partition contains ones, count them and re-use
            reuse = partition[1]
            del my_keys[-1]
            del partition[1]

        #(possibly also) reuse the smallest key of the last partition larger than 1
        smallest_key = my_keys[-1]
        new_val = partition[smallest_key]-1
        partition[smallest_key] = partition[smallest_key] - 1
        reuse += smallest_key
        if new_val == 0:
            del my_keys[-1], partition[smallest_key]

        #take the part to reuse and see how many (smallest_key-1)s we can squeeze out of it
        skmo = (smallest_key-1)
        nis, remain = divmod(reuse, skmo)
        partition[skmo] = nis
        my_keys.append(skmo)
        if remain:
            partition[remain] = 1
            my_keys.append(remain)
        bins=sum(partition.values())
        if bins<=k:
            yield partition

def GenerateaProduct(l,q,k):
    """Generate P_{ell q}^{k} as defined in Eq.(40)"""
    Product=[]
    #First enumerate all partitions of the excess q into l pieces
    PqSet=Partitions(q,l)
    for Pq in PqSet:
        LPq=sum(Pq.values())
        #make new dict with shifted values, then include 
        a={}
        for f in Pq:
            a[f+1]=Pq[f]
        #if the length of the partition is less than l, append some ones
        if (l-LPq)!=0:
            a[1]=l-LPq
        elem={'weight':MultiplicityFactor(l,Pq)*(k+1-l-q),'powers':a}
        Product.append(elem)
    return Product
    
def GenerateEtaLHSs(k):
    """Generate all values of \eta_{mk} as defined in Eq.(39), m=1,...,k, as a list"""
    etas=[]
    for m in range(1,k+1):
        firstcase={'weight' : (k+1-m), 'powers': {1 : m}}
        etaLHS=[]
        etaLHS.append(firstcase)
        for q in range(1,k-m+1):
            etaLHS+=GenerateaProduct(m,q,k)
        etas.append(etaLHS)
    return etas

def MultiplicityFactor(l,Pq):
    """Compute the MultiplicityFactor of a given integer partition Pq for a multinomial of degree l."""
    LPq=sum(Pq.values())
    #Compute log of falling factorial l_{LPq}
    numer = 0.0
    for i in range(LPq):
        numer+=log(l-i)
    denom=0.0
    for term in Pq:
        denom+=log(factorial(Pq[term]))
    fac=exp(numer-denom)
    return fac

def GenerateEtaRHSs(k):
    """Generate all numerical values of \eta_{mk}, as defined in Eq.(30), m=1,...,k, as a list"""
    etas=[]
    for m in range(1,k+1):
        eta=0
        for j in range(m+1):
            xi_jk = (k+1)*((j+1)**k-1)
            eta+=((-1)**(j+m))*xi_jk*comb(m,j)
        etas.append(eta)
    return etas

def GenerateNumericalproduct(term,a):
    """Given a monomial in a and the numerical values of a, return the numerical value"""
    k=len(a)
    val=term['weight']
    for x in term['powers']:
        val*=a[k-x]**(term['powers'][x])
    return val

def Solveeqns(etaLHS,etaRHS):
    """Solve the eta equations for the vector a"""
    k=len(etaLHS)
    #Start at the top and work down
    a=[0]*k
    #Top equation is always a_{k-1}^{k}=(k+1)!
    tmp=log(factorial(k+1))
    a[k-1]=exp(tmp/(k*1.0))
    for p in range(1,k):
        sought=p+1
        rhs=etaRHS[k-p-1]
        for term in etaLHS[k-p-1]:
            if sought in term['powers']:
                soughtterm=term
            else:
                rhs-=GenerateNumericalproduct(term,a)
        #Everything except the (single) term containing a[k-p] has been moved the the rhs
        #Peel off any other factors of a
        val=soughtterm['weight']
        for x in soughtterm['powers']:
            if x!=sought:
                val*=a[k-x]**(soughtterm['powers'][x])
        a[k-sought]=rhs/val
    return a

def PrintEtaEquations(LHS, RHS,k):
    """Print out the contents of the LHS and RHS dicts as a human-readable equation"""
    for p in range(k):
        print 'm=',k-p
        mystr = ''
        for term in etaLHS[k-p-1]:
            powstr = ''
            for pow in term['powers']:
                powstr += 'a['+str(pow)+']**'+str(term['powers'][pow])+' *'
            mystr+=str(term['weight'])+' * '+powstr[:-2] +' + '
        mystr = mystr[:-2]+' = '+str(etaRHS[k-p-1])
        print mystr

if __name__ == '__main__':
    for k in range(1,7):
        print 'k',k
        etaRHS=GenerateEtaRHSs(k)
        etaLHS=GenerateEtaLHSs(k)
        PrintEtaEquations(etaLHS, etaRHS,k)
        a=Solveeqns(etaLHS,etaRHS)
        print 'a:', a
        print '\n\n'
\end{python}

\bibliographystyle{prsty}
\bibliography{running}

\end{document}